\documentclass[preprint,1p]{elsarticle}
\usepackage{amsmath,amsfonts,amssymb,graphics,graphicx,epsfig,color,times}
\usepackage{graphics}
\usepackage[utf8x]{inputenc}
\usepackage{verbatim}
\usepackage{color}
\usepackage{bbm} 
\usepackage{subfigure}
\usepackage{hyperref}
\usepackage{mathrsfs}
\journal{Annals of Physics}
\begin{document}
\begin{frontmatter}
\title{Superradiant cascade emissions in an atomic ensemble via four-wave mixing}
\author{H. H. Jen}
\ead{sappyjen@gmail.com}
\address{Institute of Physics, Academia Sinica, Taipei 11529, Taiwan, R. O. C.}

\begin{abstract}
We investigate superradiant cascade emissions from an atomic ensemble driven by two-color classical fields.\ The correlated pair of photons (signal and idler) is generated by adiabatically driving the system with large-detuned light fields via four-wave mixing.\ The signal photon from the upper transition of the diamond-type atomic levels is followed by the idler one which can be superradiant due to light-induced dipole-dipole interactions.\ We then calculate the cooperative Lamb shift (CLS) of the idler photon, which is a cumulative effect of interaction energy.\ We study its dependence on a cylindrical geometry, a conventional setup in cold atom experiments, and estimate the maximum CLS which can be significant and observable.\ Manipulating the CLS of cascade emissions enables frequency qubits that provide alternative robust elements in quantum network.
\end{abstract}
\end{frontmatter}

\section{Introduction}
Superradiance \cite{Dicke1954} is a coherent and collective radiation from a multi-atom system.\ The radiation intensity is proportional to the square of particle number along with a shorter time variation, which conserves radiation energy.\ This collective radiation originates from a common light-matter interaction, through which spontaneously emitted photons can be reabsorbed by the atoms \cite{Stroud1972, Gross1982} if they are close to each other, and interatomic phase correlations build up \cite{Menshikov1999}.\ This spontaneous emission decays in an enhanced rate proportional to the number of particles, which is very different from spontaneous emissions of independent and uncorrelated single atoms \cite{Mandel1995}. 

Many theoretical works from different perspectives have investigated the superradiant emission.\ The microscopic mechanism of superradiance is shown to be related to induced dipole-dipole interactions \cite{Stephen1964, Lehmberg1970} between two atoms, in which spontaneous decay rate and frequency shift \cite{Arecchi1970, Morawitz1973} depend on their spatial separations.\ The canonical review articles on the frequency shift including cooperative Lamb shift (CLS) and Lorentz-Lorenz shift from a local field correction in the extended medium of a slab or a sphere can be found in the references of \cite{Friedberg1973, Manassah1983}.\ A master equation is formulated \cite{Agarwal1970, Bonifacio1971} to study the statistical property of superradiance \cite{Bonifacio1971}, and a diagonalization of coupling matrix \cite{Ernst1968, Ressayre1977} reveals the essential information of superradiance decay constants.\ One of the characteristics in superradiance is its directionality of emission \cite{Ernst1968, Rehler1971}, and there is also a semiclassical treatment incorporating propagation effect, which indicates a threshold condition of cooperative emission in an extended medium \cite{Bonifacio1975, MacGillivray1976}.\ An alternative approach of quantum trajectory method \cite{Carmichael2000, Clemens2003} for superradiance tries to unravel the physics of successive sub- and superradiant photons, and positive-P phase-space method simulations show an enhanced decay in the second order correlation function by including quantum fluctuations in a diamond-type atomic system \cite{Jen2012}.

The preparation of single-photon absorption in the atomic ensemble followed by a collective \cite{Eberly2006} and directional emission \cite{Scully2006} raises the interests of single-photon superradiance.\ This singly-inverted system reduces Dicke's full eigenstates of spin-$1/2$ system into only N state bases, and shows dynamical evolutions \cite{Mazets2007, Svidzinsky2008, Svidzinsky2008-2, Friedberg2008, Friedberg2008-3} from a specified initial state.\ The initially symmetrical excited state for two-level atoms radiates superradiantly while other less symmetrical excited states radiate in a slower rate with smaller probabilities \cite{Mazets2007}.\ Significant collective Lamb shift in a spherical geometry is also calculated in such singly-inverted system \cite{Scully2009}.\ To correctly account for the cooperative effects in a dense medium, it is crucial to include the counter rotating-wave-approximation (RWA) terms \cite{Friedberg2008} rather than excluding them as incomplete treatments \cite{Scully2006, Mazets2007, Svidzinsky2008, Svidzinsky2008-2}.\ Recently some experiments demonstrate superradiance or CLS in a variety of atomic systems, including atoms in a planar cavity \cite{Rohlsberger2010}, atomic vapor layers \cite{Keaveney2012}, cold atoms \cite{Pellegrino2014}, and cascade atomic systems \cite{Chaneliere2006, Srivathsan2013}.

The cascade atomic system provides a source for telecommunication bandwidth in its upper transition \cite{Chaneliere2006}.\ This correlated photon pair can realize long-distance quantum communication \cite{Radnaev2010, Jen2012-2} enabling a low-loss quantum repeater \cite{Briegel1998} in the DLCZ (Duan-Lukin-Cirac-Zoller) protocols \cite{Duan2001}.\ In this paper we investigate the superradiance of correlated photon pair from a diamond-type atomic ensemble driven by two-color classical fields, and calculate the CLS in an optically-thick cylindrical atomic system.\ We solve for Schr\"{o}dinger equations for such atomic system, and derive two-photon state function in section 2.\ Single atomic excitation adiabatically follows two excitation fields, and subsequently decays through cascade transitions via four-wave mixing (FWM).\ Counter-RWA terms in the Hamiltonian are required to correctly deduce the CLS which is a cumulative dipole-dipole interaction energy.\ In section 3 we calculate and study the CLS of two-photon state in a cylindrical geometry, a conventional setup in cold atom experiments.\ We conclude and summarize in section 4, and the inclusion of virtual photons exchange and adiabatic approximation of excitation process are detailed in Appendices A and B respectively.
\section{Two-photon state function}
\subsection{Hamiltonian and Schr\"{o}dinger equation of motion}
\begin{figure}[t]
\centering
\includegraphics[width=10cm,height=5.5cm]{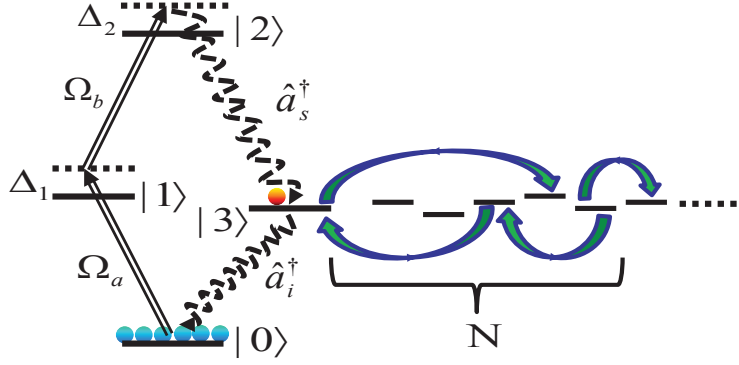}
\caption{(Color online) Four-level atomic system driven by two classical fields ($\Omega_{a,b}$) and spontaneously emitted photon pair ($\hat{a}^\dagger_{s,i}$).\ Single and two-photon detunings are $\Delta_{1,2}$.\ Schematic collective couplings between N state bases are denoted by arrows, and slightly shifted energy levels (not to scale) are due to light-induced dipole-dipole interaction energy.}\label{fig1}
\end{figure}

We consider an ensemble of N four-level atoms excited by two classical fields, and subsequently the signal and idler photons are spontaneously emitted as shown in Fig. \ref{fig1}.\ Without loss of generality, we assume that these identical atoms distribute randomly in a cylindrical geometry with a uniform density as in conventional cold atom experiments \cite{Chaneliere2006}.\ We use dipole approximation of light-matter interaction and rotating wave approximation (RWA), and the Hamiltonian in interaction picture reads
\begin{align}
V_{\rm I}&=-\hbar\Delta_1\sum^N_{\mu=1} |1\rangle_\mu\langle 1|-\hbar\Delta_2\sum^N_{\mu=1} |2\rangle_\mu\langle 2| -\frac{\hbar}{2}\Big\{\Omega_a\hat{P}^\dag_{\mathbf{k}_a}+ \Omega_b \hat{P}^\dag_{\mathbf{k}_b}+{\rm h.c.} \Big\}\nonumber\\
&-i\hbar\Big\{\sum_{\mathbf{k}_s,\lambda_s}g_{\mathbf{k}_s}
(\epsilon_{\mathbf{k}_s,\lambda_s}\cdot\hat{d}_s^*)\hat{a}_{\mathbf{k}_s,\lambda_s}\hat{S}^\dag_{\mathbf{k}_s}e^{-i(w_{\mathbf{k}_s}-\omega_{23}-\Delta_2)t}\nonumber\\&+\sum_{\mathbf{k}_i,\lambda_i}g_{\mathbf{k}_i}
(\epsilon_{\mathbf{k}_i,\lambda_i}\cdot\hat{d}_i^*)\hat{a}_{\mathbf{k}_i,\lambda_i}\hat{I}^\dag_{\mathbf{k}_i}e^{-i(\omega_{\mathbf{k}_i}-\omega_3)t}-{\rm h.c.}\Big\}\label{H},
\end{align}
where h.c. denotes Hermitian conjugate.\ Note that the RWA Hamiltonian is valid for deriving state functions to calculate transition probabilities while non-RWA or counter-RWA terms are required to account for a complete derivation of CLS (see in Appendix A).\ The collective dipole operators are defined as
\begin{align}
\hat{P}^\dag_{\mathbf{k}_a}&\equiv\sum_{\mu}|1\rangle_\mu\langle 0|e^{i\mathbf{k}_a\cdot\mathbf{r}_\mu},~
\hat{P}^\dag_{\mathbf{k}_b}\equiv\sum_{\mu}|2\rangle_\mu\langle 1|e^{i\mathbf{k}_b\cdot\mathbf{r}_\mu},\nonumber\\
\hat{S}^\dag_{\mathbf{k}_s}&\equiv\sum_{\mu}|2\rangle_\mu\langle 3| e^{i\mathbf{k}_s\cdot\mathbf{r}_\mu},~
\hat{I}^\dag_{\mathbf{k}_i}\equiv\sum_{\mu}|3\rangle_\mu\langle 0| e^{i\mathbf{k}_i\cdot\mathbf{r}_\mu},\label{op}
\end{align}
where single photon detuning $\Delta_1$ $=$ $\omega_a$ $-$ $\omega_1$, two-photon detuning $\Delta_2$ $=$ $\omega_a$ $+$ $\omega_b$ $-$ $\omega_2$, and transition frequency $\omega_{23}$ $=$ $\omega_2$ $-$ $\omega_3$.\ Driving Rabi frequencies are $\Omega_a$ $\equiv$ $(1||\hat{d}||0)\mathcal{E}(k_a)/\hbar$, $\Omega_b$ $\equiv$ $(2||\hat{d}||1)\mathcal{E}(k_b)/\hbar$, and coupling coefficients are $g_{\mathbf{k}_s}$ $\equiv$ $(3||\hat{d}||2)\mathcal{E}(\mathbf{k}_s)/\hbar$, $g_{\mathbf{k}_i}$ $\equiv$ $(0||\hat{d}||3)\mathcal{E}(\mathbf{k}_i)/\hbar$.\ The double matrix element of the dipole moment $\hat{d}$ is independent of the hyperfine structure, and $\mathcal{E}(k)$ $=$ $\sqrt{\hbar kc/(2\epsilon_0 V)}$ where $V$ is the quantization volume.\ Polarizations of signal and idler fields are
$\epsilon_{\mathbf{k}_{s,i},\lambda_{s,i}}$, and the unit directions of dipole operators are $\hat{d}_{s,i}$ respectively.

In the limit of large detuned and weak driving fields, $\Delta_1$ $\gg$ $\sqrt{N}|\Omega_a|$, we consider only single excitation, and neglect spontaneous decay during excitations ($\Delta_1$ $\gg$ $\Gamma_1$, $\Delta_2$ $\gg$ $\Gamma_2$).\ We may express the state function as
\begin{align}
|\psi(t)\rangle&=\mathcal{E}(t)|0,{\rm vac}\rangle+\sum^N_{\mu=1} A_\mu(t)|1_\mu,{\rm vac}\rangle+\sum^N_{\mu=1} B_\mu(t)|2_\mu,{\rm vac}\rangle+\sum^N_{\mu=1}\sum_{s}C^\mu_s(t)|3_\mu,1_{\mathbf{k}_s,\lambda_s}\rangle \nonumber\\
&+\sum_{s,i}D_{s,i}(t)|0,1_{\mathbf{k}_s,\lambda_s},1_{\mathbf{k}_i,\lambda_i}\rangle,
\end{align}
where the indices $s=(\mathbf{k}_s,\lambda_s)$, $i=(\mathbf{k}_i,\lambda_i)$ for light fields, $|m_\mu\rangle$ $\equiv$ $|m\rangle_\mu|0\rangle^{\otimes N-1}_{\nu\neq\mu}$ for atomic levels $m=1,2,3$, and $|{\rm vac}\rangle$ is the vacuum photon state.\ These bare states make a complete basis of collective excitations for identical particles, which describe a complete cycle of a single atom following the excitation and spontaneously emitted fields.\ We apply Schr\"{o}dinger equation $i\hbar\frac{\partial}{\partial t}|\psi(t)\rangle$ $=$ $V_{\rm I}(t)|\psi(t)\rangle$, and the coupled equations of motion are
\begin{align}
i\dot{\mathcal{E}}&=-\frac{\Omega_a^*}{2} \sum_\mu e^{-i\mathbf{k}_a\cdot\mathbf{r}_\mu}A_\mu\label{eqn1},\\
i\dot{A}_\mu&=-\frac{\Omega_a}{2}e^{i\mathbf{k}_a\cdot\mathbf{r}_\mu}\mathcal{E}-\frac{\Omega_b^*}{2} e^{-i\mathbf{k}_b\cdot\mathbf{r}_\mu}B_\mu-\Delta_1 A_\mu\label{eqn2},\\
i\dot{B}_\mu&=-\frac{\Omega_b}{2}e^{i\mathbf{k}_b\cdot\mathbf{r}_\mu}A_\mu-\Delta_2 B_\mu-i\sum_{s}g_{\mathbf{k}_s}(\epsilon_{\mathbf{k}_s,\lambda_s}\cdot\hat{d}_s^*) e^{i\mathbf{k}_s\cdot\mathbf{r}_\mu}e^{-i(\omega_{ks}-\omega_{23}-\Delta_2)t}C^\mu_{s}\label{eqn3},\\
\dot{C}^\mu_{s}&=ig_{\mathbf{k}_s}^*(\epsilon^*_{\mathbf{k}_s,\lambda_s}\cdot\hat{d}_s)e^{-i\mathbf{k}_s\cdot\mathbf{r}_\mu}e^{i(\omega_{\mathbf{k}_s}-\omega_{23}-\Delta_2)t}B_\mu\nonumber\\
&-i\sum_{i}g_{\mathbf{k}_i}(\epsilon_{k_i,\lambda_i}\cdot\hat{d}_i^*) e^{i\mathbf{k}_i\cdot\mathbf{r}_\mu}e^{-i(\omega_{\mathbf{k}_i}-\omega_3)t}D_{s,i}\label{eqn4},\\
i\dot{D}_{s,i}&=ig_{\mathbf{k}_i}^*(\epsilon^*_{k_i,\lambda_i}\cdot\hat{d}_i)\sum_\mu e^{-i\mathbf{k}_i\cdot\mathbf{r}_\mu}e^{i(\omega_{\mathbf{k}_i}-\omega_3)t}C^\mu_{s}\label{eqn5}.
\end{align}

In the next subsection, we proceed to solve for two-photon state function from the above equation of motion in the bare state basis.
\subsection{Two-photon state in bare state basis}
Firstly we solve for signal photon state function by substituting Eq. (\ref{eqn5}) into Eq. (\ref{eqn4}), and the time derivative of $C^\mu_{s}(t)$ becomes
\begin{align}
\dot{C}^\mu_{s}(t)&=g_s^*(\epsilon^*_{s}\cdot\hat{d}_s)e^{-i\mathbf{k}_s\cdot\mathbf{r}_\mu}e^{i(\omega_{\mathbf{k}_s}-\omega_{23}-\Delta_2)t}B_\mu(t)
-\sum_\nu\sum_{\mathbf{k}_i,\lambda_i}|g_i|^2|\epsilon_{\mathbf{k}_i,\lambda_i}\cdot\hat{d}_i^*|^2e^{i\mathbf{k}_i\cdot(\mathbf{r}_\mu-\mathbf{r}_\nu)}\nonumber\\
&\times\int_{0}^t dt'e^{i(\omega_{\mathbf{k}_i}-\omega_3)(t'-t)}C^\nu_{s}(t').
\end{align}

We use Weisskopf-Wigner approach \cite{QO:Scully} which assumes a weak-coupling limit and implicitly a Markov approximation $\omega_3$ $\gg$ $\Gamma_3$ with an intrinsic relaxation time set by $1/\Gamma_3$ (see below).\ In addition the secular approximation suggests an ensemble size ($L_s$) limit of $L_s/c$ $\ll$ $1/\Gamma_3$ for the approach to be valid \cite{Cohen-Tannoudji1992, Lehmberg1970}.\ The summation of field modes in the above leads to the frequency shift and decay rate which are
\begin{align}
&\sum_\nu\sum_{\mathbf{k}_i,\lambda_i}|g_i|^2|\epsilon_{\mathbf{k}_i,\lambda_i}\cdot\hat{d}_i^*|^2e^{i\mathbf{k}_i\cdot(\mathbf{r}_\mu-\mathbf{r}_\nu)}
\int_{0}^t dt'e^{i(\omega_{\mathbf{k}_i}-\omega_3)(t'-t)}C^\nu_{s}(t'),\nonumber\\
&=\sum_\nu\int_0^\infty d\omega_iF_{\mu\nu}(k_i r_{\mu\nu})\frac{1}{2\pi}\frac{|d_i|^2\omega_i^3}{3\pi\hbar\epsilon_0c^3}C^\nu_{s}(t)
\left[\pi\delta(\omega_i-\omega_3)-i{\rm P.V.}(\omega_i-\omega_3)^{-1}\right],\nonumber\\
&=\sum_\nu C^\nu_{s}(t)\left[\frac{\Gamma_3}{2}F_{\mu\nu}(\xi)-i\Omega_{\mu\nu}^-(\xi)\right],\label{eqn_F}
\end{align}
where $\xi=k_3 r_{\mu\nu}$, and $r_{\mu\nu}=|\mathbf{r}_\mu-\mathbf{r}_\nu|$.\ $\Omega_{\mu\nu}^-(\xi)$ involves the Lamb shift ($\nu=\mu$) and collective radiation shift from dipole-dipole interaction ($\mu\neq\nu$) which we describe in details in Appendix A.\ The other frequency shift of $\Omega_{\mu\nu}^+(\xi)$ can be derived by including non-RWA terms as shown in Appendix A.

Therefore the differential equation of probability amplitude of $C^\mu_{s}(t)$ becomes
\begin{align}
\dot{C}^\mu_{s}(t)&=g_s^*(\epsilon^*_{\mathbf{k}_s,\lambda_s}\cdot\hat{d}_s)e^{-i\mathbf{k}_s\cdot\mathbf{r}_\mu}e^{i(\omega_{\mathbf{k}_s}-\omega_{23}-\Delta_2)t}B_\mu(t)\nonumber\\
&-\frac{\Gamma_3}{2}\bigg[\sum_\nu C^\nu_{s}(t)F_{\mu\nu}(\xi)+i\sum_{\nu\neq\mu}2C^\nu_{s}(t)G_{\mu\nu}(\xi)+iC^\mu_{s}(t)\bar{\Omega}\bigg],\label{eqnC}
\end{align}
where $\bar{\Omega}\equiv\Omega/(\Gamma_3/2)$, and $\Omega$ is Lamb shift.\ $F_{\mu\nu}$ and $G_{\mu\nu}$ represent the collective contributions to spontaneous decay rate and frequency shift, which are induced from the couplings between N state bases as shown schematically in Fig. \ref{fig1}.

The differential equation of Eq. (\ref{eqnC}) can be expressed in a matrix form as an eigenvalue problem, and we can solve for $C^\mu_{s}(t)$ by a similarity transformations $\hat{U}$ $=$ $(u_1,u_2,...,u_N)$ comprised of eigenvectors $u_i$,
\begin{align}
C_{s}^\mu(t)&=g_{s}^*(\epsilon^*_{\mathbf{k}_s,\lambda_s}\cdot\hat{d}_s)\int^tdt'e^{i(\omega_{\mathbf{k}_s}-\omega_{23}-\Delta_2)t'}
\sum_{l,m}\hat{U}_{\mu l}e^{\lambda_l(t-t')}\hat{U}^{-1}_{lm}B_m(t')e^{-i\mathbf{k}_s\cdot\mathbf{r}_m},\label{signal_C}
\end{align}
where $\lambda_i$ and $u_i$ are eigenvalues and eigenvectors respectively of N$\times$N matrix $M$,
\begin{align}
&M\equiv-\frac{\Gamma_3}{2}\left( \begin{array}{cccc}
1&F_{12}+i2G_{12}&...&F_{1N}+i2G_{1N} \\
F_{12}+i2G_{12}&1& ...&\vdots \\
\vdots & \vdots &\ddots &\vdots\\
F_{1N}+i2G_{1N}&...&...&1
\end{array} \right).
\end{align}

Note that we have absorbed Lamb shift into the optical frequency $\omega_3$, and the properties $F_{\alpha\beta}$ $=$ $F_{\beta\alpha}$, $G_{\alpha\beta}$ $=$ $G_{\beta\alpha}$ have been used.\ The diagonal element demonstrates a single atomic spontaneous decay, and off-diagonal elements represent the couplings between atoms.\ The cross couplings are in similar order of magnitude indicating highly dynamical interactions in the system. 

We then substitute Eq. (\ref{signal_C}) into Eq. (\ref{eqn5}), and derive the two-photon state function in the bare state basis
\begin{align}
D_{s,i}(t)&=g_{i}^*g_{s}^*(\epsilon^*_{\mathbf{k}_i,\lambda_i}\cdot\hat{d}_i)(\epsilon^*_{\mathbf{k}_s,\lambda_s}\cdot\hat{d}_s)\int^t\int^{t'}dt''dt'
e^{i(\omega_{\mathbf{k}_i}-\omega_3)t'}e^{i(\omega_{\mathbf{k}_s}-\omega_{23}-\Delta_2)t''}
\nonumber\\&\times\sum_{\mu,l,m}e^{-i\mathbf{k}_i\cdot\mathbf{r}_\mu}\hat{U}_{\mu l}e^{\lambda_l(t'-t'')}\hat{U}^{-1}_{lm}B_m(t'')e^{-i\mathbf{k}_s\cdot\mathbf{r}_m}.
\end{align}
In the adiabatic approximation of the excitation process described in Appendix B, we may derive $B_m(t)$, and the single and two-photon probability amplitudes become
\begin{align}
C_{s}^\mu(t)&=g_{s}^*(\epsilon^*_{\mathbf{k}_s,\lambda_s}\cdot\hat{d}_s)\int^tdt'e^{i(\omega_{\mathbf{k}_s}-\omega_{23}-\Delta_2)t'}b(t')\sum_{l,m}\hat{U}_{\mu l}e^{\lambda_l(t-t')}\hat{U}^{-1}_{lm}e^{i(\mathbf{k}_a+\mathbf{k}_b-\mathbf{k}_s)\cdot\mathbf{r}_m},\\
D_{s,i}(t)&=g_{i}^*g_{s}^*(\epsilon^*_{\mathbf{k}_i,\lambda_i}\cdot\hat{d}_i)(\epsilon^*_{\mathbf{k}_s,\lambda_s}\cdot\hat{d}_s)\int^t\int^{t'}dt''dt'
e^{i(\omega_{\mathbf{k}_i}-\omega_3)t'}e^{i(\omega_{ks}-\omega_{23}-\Delta_2)t''}b(t'')
\nonumber\\&\times\sum_{\mu,l,m}e^{-i\mathbf{k}_i\cdot\mathbf{r}_\mu}\hat{U}_{\mu l}e^{\lambda_l(t'-t'')}\hat{U}^{-1}_{lm}e^{i(\mathbf{k}_a+\mathbf{k}_b-\mathbf{k}_s)\cdot\mathbf{r}_m} \label{two_bare},
\end{align}
where the time evolution of two-photon state incorporates two-time integration of pulse shapes $b(t)$ $=$ $\Omega_a(t)\Omega_b(t)/(4\Delta_1\Delta_2)$ which results from adiabatic driving process (See Appendix B).

The cascade two-photon state is expressed of discrete sum of N eigenvalues $\lambda_l$ shown in the time evolution and sum of eigenvectors sandwiched by phase factors induced by four fields interacting with the system.\ In the above treatment where we use the bare state basis, the signal spontaneous emission comes from the adiabatic transfer of detuned two-color driving fields, and the idler spontaneous emission comes from N intermediate excited states.\ These solutions of probability amplitudes demonstrate a dynamical coupling between N atoms, and two-photon state is generated from all possible excitations of one of the atomic ensemble.\ In the next section, we rotate the coupling matrix $M$ to the new basis, and we find the superradiant idler photon with its decay constant coming from one of the eigenvalues.

The superradiance ($\lambda_l$ $>$ $\Gamma_3$) and subradiance ($\lambda_l$ $<$ $\Gamma_3$) are embedded in these eigenvalues $\lambda_l$ which depend on the density and geometry of atomic ensemble.\ For a low density where $r_{\alpha\beta}$ $\gg$ $1/k_3$, $F_{\alpha\beta}$ $\approx$ $0$, 
$G_{\alpha\beta}$ $\approx$ $0$, and the coupling matrix $M$ becomes an identity matrix times an overall constant $-\Gamma_3/2$.\ The transformation matrix
$\hat{U}$ is then also an identity matrix.\ Therefore the two-photon state from N diamond-type atoms behaves as no difference from a single atom.\ On the contrary in Dicke's limit where the dimensions of ensemble are smaller than radiation wavelength, $F_{\alpha\beta}$ $\approx$ $1$, and $G_{\alpha\beta}$ becomes divergent when $r_{\alpha\beta}$ $\rightarrow$ $0$.\ Without considering this divergent CLS, one of the eigenvalues becomes $\textrm{Re}(\lambda_1)$ $=$ $-N\Gamma_3/2$ proportional to the number of particles, reminiscent of Dicke's state $|l=N/2,m=1-N/2\rangle$ \cite{Dicke1954, Mandel1995} where $m$ is the half difference of atomic populations in the excited and ground states.

In the next section, we construct the basis in terms of the symmetrical and N-1 unsymmetrical states.\ In this basis we are able to investigate the superradiance and CLS of the two-photon state.
\subsection{Two-photon state in phased symmetrical state}
Here we investigate the collective radiation decay and energy shift of two-photon state in the basis constructed from the phased symmetrical state \cite{Mazets2007, Svidzinsky2008} which provides a preference for the system in the limit of large number of atoms.\ Consider a unitary transformation $\hat{S}$ where a phase factor in an extended ensemble is introduced into basis \cite{Mazets2007},
\begin{align}
|\phi_l\rangle&=\sum_\mu S_{l\mu}|3\rangle_\mu\otimes|0\rangle_{\lambda\neq\mu}^{\otimes N-1},\\
S_{l\mu}&=e^{i\bar{\mathbf{k}}\cdot\mathbf{r}_\mu}f_{l\mu},\\
f_{l\mu}&\equiv\frac{1}{\sqrt{N}}\delta_{lN}+\Bigg[\left(-\frac{1}{\sqrt{N}}-\frac{1+1/\sqrt{N}}{N-1}\right)\delta_{\mu N} +\left(\frac{1+1/\sqrt{N}}{N-1}-\delta_{\mu l}\right)\Bigg]\delta_{l\neq N},
\end{align}
where $\bar{\mathbf{k}}$ $=$ $\mathbf{k}_a$ $+$ $\mathbf{k}_b$ $-$ $\mathbf{k}_s$, and the extra phase factor introduced above is from the observation in Eq. (\ref{two_bare}).\ 

The phased symmetrical state in the new basis is
\begin{align}
|\phi_N\rangle=\frac{1}{\sqrt{N}}\sum_{\mu}e^{i\bar{\mathbf{k}}\cdot\mathbf{r}_\mu}|3\rangle_\mu\otimes|0\rangle_{\lambda\neq\mu}^{\otimes N-1},
\end{align}
and the other $N-1$ unsymmetrical states are
\begin{align}
|\phi_l\rangle &=-\frac{e^{i\bar{\mathbf{k}}\cdot\mathbf{r}_N}}{\sqrt{N}}|3\rangle_N|0\rangle^{\otimes N-1}_{\lambda\neq N}
+\sum_{j=1}^{N-1}\left(\frac{1+1/\sqrt{N}}{N-1}-\delta_{jl}\right)e^{i\bar{\mathbf{k}}\cdot\mathbf{r}_j}|3\rangle_j|0\rangle^{\otimes N-1}_{\lambda\neq j},
\end{align}
where $l=1,2,...,N-1$, and they are orthogonal to each other and normalized to one.

In this new rotational basis, the signal state function has the identity
\begin{align}
\sum^N_{\mu=1}\sum_{k_s,\lambda_s}C^\mu_s(t)|3_\mu,1_{\mathbf{k}_s,\lambda_s}\rangle=\sum^N_{l=1}\sum_{k_s,\lambda_s}C^l_s(t)|\phi_l,1_{\mathbf{k}_s,\lambda_s}\rangle,\nonumber
\end{align}
where these coefficients are related by the unitary transformation matrix that $C_s^\mu=\sum_l S_{l\mu}C_s^l$ and $C_s^l=\sum_\mu S_{l\mu}^*C_s^\mu$.\ A new set of equation of motion in this basis can be derived, and we have for the equation of $C_{s}^l$,
\begin{align}
\dot{C}^l_{s}&=\sqrt{N}g_{\mathbf{k}_s}^*(\epsilon^*_{\mathbf{k}_s,\lambda_s}\cdot\hat{d}_s)b(t)e^{i(\omega_{\mathbf{k}_s}-\omega_{23}-\Delta_2)t}\delta_{lN}
-\sum_{l'}\sum_{\mu\nu}S_{l\mu}^*M_{\mu\nu}S_{l'\nu}C^{l'}_s,
\end{align}
where we have used the property of $\sum_\mu f_{k\mu}$ $=$ $\sqrt{N}\delta_{kN}$, and $M$ is the coupling matrix in bare state basis.\ The coupling matrix in this new rotational basis becomes (let $K_{\mu\nu}(\xi)$ $\equiv$ $[F_{\mu\nu}(\xi)+i2G_{\mu\nu}(\xi)]e^{-i\bar{\mathbf{k}}\cdot(\mathbf{r}_\mu-\mathbf{r}_\nu)}$)
\begin{align}
\bar{M}&=(SAS^\dagger)^T\nonumber\\
&=-\frac{\Gamma_3}{2} \left(
\begin{array}{cccc}
1+\sum_{\mu\neq\nu}f_{1\nu}f_{1\mu}K_{\mu\nu}(\xi)
&...&\sum_{\mu\neq\nu}f_{1\nu}f_{N\mu}K_{\mu\nu}(\xi) \\
\sum_{\mu\neq\nu}f_{2\nu}f_{1\mu}K_{\mu\nu}(\xi)
&\ddots&\vdots\\
\vdots&\vdots&\vdots \\
\sum_{\mu\neq\nu}f_{N\nu}f_{1\mu}K_{\mu\nu}(\xi)
&...&1+\frac{1}{N}\sum_{\mu\neq\nu}K_{\mu\nu}(\xi)
\end{array} \right),
\end{align}
where the properties $\sum_\mu f_{l\mu}$ $=$ $\sqrt{N}\delta_{lN}$, $\sum_\nu f_{l\nu}f_{l'\nu}$ $=$ $\delta_{ll'}$, $M=M^T$ are used, and $T$ means transpose.\ 
The diagonal element $\bar{A}_{ii},~i\leq N-1$ has the order of $1$, which indicates a spontaneously decayed single atom, and $\bar{A}_{NN}$ involves the superradiant decay rate and CLS of the symmetrical state.

Similarly we may diagonalize the above matrix with a similarity transformation matrix $\hat{U}$, and the solutions for probability amplitudes of single and two-photon states become
\begin{align} 
C_{s}^l(t)&=\sqrt{N}g_{s}^*(\epsilon^*_{\mathbf{k}_s,\lambda_s}\cdot\hat{d}_s)\int^tdt'e^{i(\omega_{\mathbf{k}_s}-\omega_{23}-\Delta_2)t'}b(t')
\sum_{j}\hat{U}_{lj}e^{\lambda_j(t-t')}\hat{U}^{-1}_{jN},\label{c}
\end{align}
\begin{align}
D_{s,i}(t)&=\sqrt{N}g_{i}^*g_{s}^*(\epsilon^*_{\mathbf{k}_i,\lambda_i}\cdot\hat{d}_i)(\epsilon^*_{\mathbf{k}_s,\lambda_s}\cdot\hat{d}_s)\int^t\int^{t'}dt''dt'
 e^{i(\omega_{\mathbf{k}_i}-\omega_3)t'}e^{i(\omega_{\mathbf{k}_s}-\omega_{23}-\Delta_2)t''}b(t'')\nonumber\\
&\times\sum_{\mu}e^{i\Delta\mathbf{k}\cdot\mathbf{r}_\mu}\sum^N_{l,j}f_{l\mu}\hat{U}_{lj}e^{\lambda_j(t'-t'')}\hat{U}^{-1}_{jN},\label{d}
\end{align}
where $\Delta\mathbf{k}$ $\equiv$ $\mathbf{k}_a$ $+$ $\mathbf{k}_b$ $-$ $\mathbf{k}_s$ $-$ $\mathbf{k}_i$ indicates FWM mismatch.\ The state functions show a specific dependence on $N$th column of $\hat{U}^{-1}$ which signifies the difference of the new rotational basis from bare state one.

One way to simplify the solutions is to examine the matrix elements of $\bar{A}$.  In the limit of large number of atoms, $\bar{A}_{iN}$ and $\bar{A}_{Ni}$
with $i\neq N$ are negligible in the order of $\frac{1}{\sqrt{N}}$ compared to $\bar{A}_{NN}$.  It can be shown to use a Gaussian density distribution in
continuous limit \cite{Mazets2007} with approximation of $f_l\mu\approx N^{-1}-\delta_{l\mu}$ where $l\neq N$.\ The order of magnitude specifies the coupling
strength between the phased symmetrical state and the other $N-1$ less symmetrical ones.\ One can show the cross-coupling elements of $N-1$ non-symmetrical states are not as negligible as $\bar{A}_{iN}$ or $\bar{A}_{Ni}$.\ It is estimated as less than $1/(\sqrt{\bar{A}_{NN}}N^{1/6})$ compared to $\bar{A}_{NN}$, but it does not come into play with our two-photon state due to the weak coupling constants of $\bar{A}_{iN}$ and $\bar{A}_{Ni}$.\ To lowest order of simplification of matrix $\bar{A}$, we legitimately set $N$th row and column of $\bar{A}$ to zero except for the element $\bar{A}_{NN}$ and we find that $\hat{U}_{lN}$ $=$ $\delta_{lN}$ and $\hat{U}^{-1}_{jN}$ $=$ $\delta_{jN}$ which can be regarded as decoupling from this symmetrical state $|\phi_N\rangle$.\ The two-photon state therefore becomes
\begin{align}
D_{s,i}(t)&=g_{i}^*g_{s}^*(\epsilon^*_{\mathbf{k}_i,\lambda_i}\cdot\hat{d}_i)(\epsilon^*_{\mathbf{k}_s,\lambda_s}\cdot\hat{d}_s)\sum_{\mu}e^{i\Delta\mathbf{k}\cdot\mathbf{r}_\mu}\int^t\int^{t'}dt''dt'e^{i(\omega_{\mathbf{k}_i}-\omega_3)t'}e^{i(\omega_{\mathbf{k}_s}-\omega_{23}-\Delta_2)t''}\nonumber\\
&\times b(t'')e^{\lambda_N(t'-t'')}.\label{twomu}
\end{align}

The above is one of the central results in this paper.\ The condition of FWM is embedded in the summation of four-wave phase factor $\sum_{\mu}e^{i\Delta\mathbf{k}\cdot\mathbf{r}_\mu}$ which is maximized when $\Delta\mathbf{k}$ $=$ $0$.\ The cascade emission has the most significant contribution of the phased symmetrical state where $\lambda_N$ represents the timescale of spontaneous decay and CLS for the idler transition.\ Note that the cooperative effect for this correlated photon pair appears only on the idler transition, which is due to the large-detuned FWM driving process that the atomic system is adiabatically prepared to the upper singly-excited state.\ In deriving this two-photon state function with induced dipole-dipole interactions in the idler transition, the superradiant decay rate and CLS of the idler photon have no difference from the emission of a phased symmetric state in a two-level atomic system.
\begin{figure}[t]
\centering
\includegraphics[width=12cm,height=5.5cm]{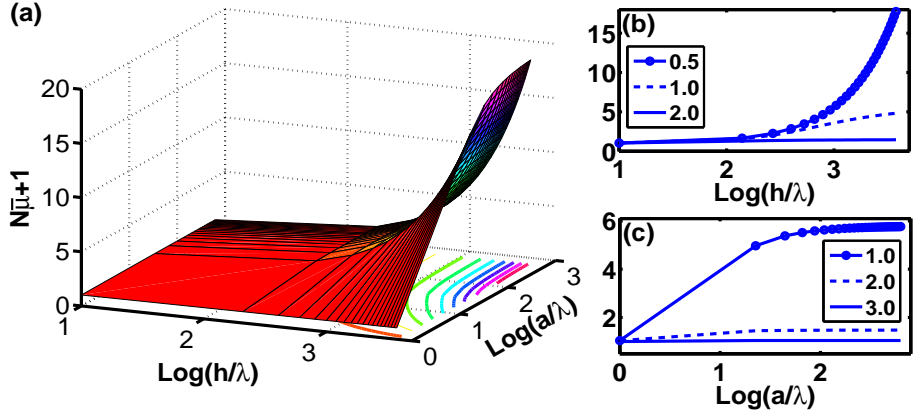}
\caption{(Color online) (a) A superradiance decay factor $N\bar{\mu}+1$'s dependence on height $h$ and radius $a$ in a cylindrical geometry.\ The enhanced superradiance factor has a stronger dependence on the height, which can be seen in the sectional plot of (b) while it saturates as radius increases as shown in (c).\ The legends in (b,c) denote the values of $\log(a/\lambda)$ and $\log(h/\lambda)$ respectively.\ D1 transition of the idler photon is chosen for Rb atoms, and the atomic density is $8\times 10^{10}$ cm$^{-3}$.}\label{fig2}
\end{figure}

The real part of $\lambda_N$ can be expressed in terms of geometrical constant \cite{Rehler1971} of a cylindrical ensemble,
\begin{align}
\textrm{Re}(\lambda_N)&=-\frac{\Gamma_3}{2}\bigg[1+\frac{1}{N}\sum_{\mu\neq\nu}F_{\mu\nu}(k_3r_{\mu\nu})e^{-i\bar{\mathbf{k}}\cdot(\mathbf{r}_\mu-\mathbf{r}_\nu)}\bigg],\nonumber\\
&=-\frac{\Gamma_3}{2}\bigg[\frac{1}{N}\sum_{\mu,\nu}F_{\mu\nu}(k_3r_{\mu\nu})e^{-i\bar{\mathbf{k}}\cdot(\mathbf{r}_\mu-\mathbf{r}_\nu)}\bigg],\nonumber\\
&=-\frac{\Gamma_3}{2} N\int d\Omega_3\frac{3}{8\pi}\left[1-(\hat{k}_3\cdot\hat{d})^2\right]\Gamma(\bar{\mathbf{k}},\mathbf{k}_3)=-\frac{\Gamma_3}{2}(N\bar{\mu}+1),\label{eqn_lambda}
\end{align}
where in the second line we use $F_{\mu\mu}$ $=$ $1$.\ $\Omega_3$ is the solid angle of idler photon with wavevector $\mathbf{k}_3$, and $\Gamma(\bar{\mathbf{k}},\mathbf{k}_3)$ $\equiv$ $\frac{1}{N^2}\sum_{\mu,\nu}$ $e^{i(\bar{\mathbf{k}}-\mathbf{k}_3)\cdot(\mathbf{r}_\mu-\mathbf{r}_\nu)}$ which is the ensemble average with randomly distributed atomic positions $\mathbf{r}_\mu$ and $\mathbf{r}_\nu$.\ In deriving Eq. (\ref{eqn_lambda}) we first substitute $F_{\mu\nu}(k_3 r_{\mu\nu})$ for $\int d\Omega_3\frac{3}{8\pi}[1-(\hat{k}_3\cdot\hat{d})^2]e^{i\mathbf{k}_3\cdot(\mathbf{r}_\mu-\mathbf{r}_\nu)}$ \cite{Lehmberg1970} as in Eq. (\ref{eqn_F}) where the property of polarizations $\sum_\lambda\epsilon_{\mathbf{k},\lambda}\epsilon_{\mathbf{k},\lambda}$ $=$ $1-\mathbf{k}\mathbf{k}/|\mathbf{k}|^2$ is used with $\mathbf{k}\mathbf{k}$ in a dyadic form \cite{QO:Scully}.\ Then we go to the continuous limit by replacing $\sum_\mu$ with $\frac{1}{V}\int d^3r$, and relate to $\bar{\mu}$ of a cylindrical geometry \cite{Rehler1971}.\ We note that the excitation (absorbing two-color classical fields with an emitting signal photon) propagates along the height of the cylinder, which is $\bar{\mathbf{k}}$.\ The geometrical constant $\bar{\mu}$ for a cylindrical ensemble (of height $h$ and radius $a$) therefore is \cite{Rehler1971}
\begin{align}
\bar{\mu}&=\frac{6(N-1)}{NA^2H^2}\int_{-1}^1\frac{dx(1+x^2)}{(1-x)^2(1-x^2)}\sin^2\left[\frac{1}{2}H(1-x)\right]J^2_1\left[A\sqrt{1-x^2}\right],
\end{align}
where $H$ $=$ $k_3 h$ and $A$ $=$ $k_3 a$ are dimensionless length scales, and circular polarizations are considered \cite{Rehler1971}.\ $J_1$ is the Bessel function of the first kind.\ As shown in Fig.(\ref{fig2}), we demonstrate how superradiance decay depends on length and radius of a cylindrical ensemble.

The real part retrieves Dicke's result when atoms are confined within the radiating wavelength and $\bar{\mu}=1-1/N$.\ When the average separation of the ensemble is much larger than emission wavelength, dipole-dipole interaction becomes negligible, and we have single atomic spontaneous emission $\bar{\mu}\rightarrow 0$.\ The frequency shift of this symmetrical state can be significant if atoms are close to each other.\ It is a cumulative contribution from dipole-dipole interactions which are oscillatory in space as shown in Appendix A.
\section{Cooperative Lamb shift of the phased symmetrical state}
In this section, we calculate the cooperative Lamb shift of phased symmetrical state, which is 
\begin{align}
\textrm{Im}(\lambda_N)&=-\frac{\Gamma_3}{2}\frac{2}{N}\sum_{\mu\neq\nu}G_{\mu\nu}(k_3r_{\mu\nu})e^{-i\bar{\mathbf{k}}\cdot(\mathbf{r}_\mu-\mathbf{r}_\nu)},\nonumber\\
&=\frac{\Gamma_3}{Nk_3^3}\textrm{P.V.}\int_{-\infty}^\infty\frac{dk}{2\pi}\frac{k^3}{k-k_3}\sum_{\mu\neq\nu}F_{\mu\nu}(kr_{\mu\nu})
e^{-i\bar{\mathbf{k}}\cdot(\mathbf{r}_\mu-\mathbf{r}_\nu)},\nonumber\\
&=\frac{\Gamma_3}{k_3^3}\textrm{P.V.}\int_{-\infty}^\infty\frac{dk}{2\pi}\frac{k^3}{k-k_3}N\bar{\mu}(k),\label{f1}
\end{align}
where P.V. denotes principal value, and we note that geometrical constant $\bar{\mu}(k)$ has a momentum dependence in the integral.\ Interestingly the CLS ${\rm Im}(\lambda_N)$ relates to the integral of spontaneous decay rate proportional to $N\bar{\mu}(k)$, which is a Hilbert transform \cite{Cohen-Tannoudji1992} if we put back the Lamb shift into our expression.\ The Lamb shift is an intrinsic level shift induced by vacuum fluctuation while the CLS depends on the size and geometry of the atomic ensemble indicating a very different physical mechanism.
\begin{figure}[t]
\centering
\includegraphics[width=12cm,height=5cm]{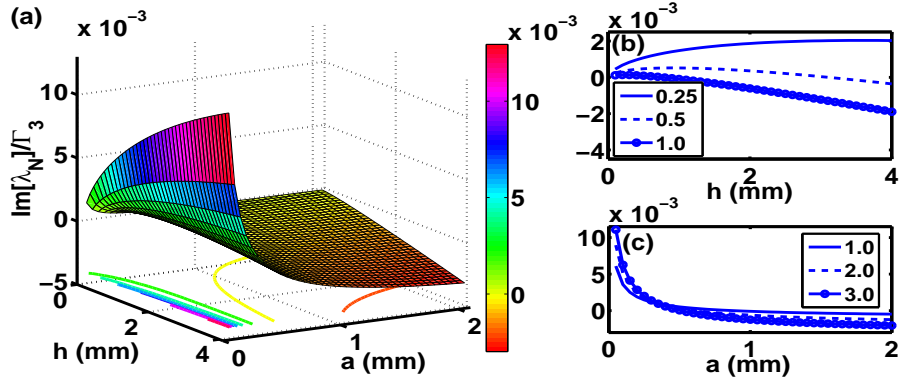}
\caption{(Color online) Cooperative Lamb shift (CLS) in a cylindrical geometry.\ (a) 3D plot of CLS on height $h$ and radius $a$, and (b,c) cross sections on $h$ and $a$ respectively with the values of $a$ and $h$ indicated in the legends.\ The atomic density and other parameters are the same as in Fig \ref{fig2}.\ Short radius behavior involves a sharp increase which can be seen in Fig. \ref{fig4}(b).}\label{fig3}
\end{figure}

To calculate the integral, we may start from the analytic form of $\mu(k)$, and re-express Eq. (\ref{f1}) by renormalizing $k$ $\rightarrow$ $k_3 k$, 
\begin{align}
\textrm{Im}(\lambda_N)&=\Gamma_3\textrm{P.V.}\int_{-\infty}^\infty\frac{dk}{2\pi}\frac{k^3}{k-1}\frac{6(N-1)}{k^4A^2H^2}\nonumber\\
&\times\int_{-1}^1(1+x^2)
\frac{\sin^2\left[\frac{1}{2}kH(1-x)\right]J^2_1\left[kA\sqrt{1-x^2}\right]}{(1-x)^2(1-x^2)}dx.
\end{align}
The function of integral over $x$ changes slowly compared to the change of $k$, and we may move it outside of the integral similar to the nonperturbative treatment in the radiation level shift \cite{Cohen-Tannoudji1992}.\ Taking $k$ $=$ $1$ for the integral of $x$ where the integral of $k$ is most appreciable, we have
\begin{align}
\textrm{Im}(\lambda_N)&=N\bar{\mu}(k_3)\Gamma_3\times\textrm{P.V.}\int_{k_m}^{k_M} \frac{dk}{k}\left(\frac{1}{k-1}+\frac{1}{k+1}\right),\label{cutoff}
\end{align}
where we introduce the infrared ($k_m$) and ultraviolet ($k_M$) energy cutoffs \cite{Cohen-Tannoudji1992, Donaire2012} to the integral, which represent the lowest and largest energy scales in the system.\ Then we have the CLS in the leading order of $k_m/k_3$ and $k_3/k_M$,
\begin{align}
\textrm{Im}(\lambda_N)&=2N\bar{\mu}(k_3)\Gamma_3\left(\frac{k_m}{k_3}-\frac{k_3}{k_M}\right),\label{shift}
\end{align}
which indicates a redshift \cite{Scully2009, Keaveney2012, Pellegrino2014} for the level $|3\rangle$ if we let $k_M\rightarrow\infty$.\ In Fig. \ref{fig3}, CLS shows a geometrical dependence on height and radius in a cylindrical atomic ensemble.\ We use $k_m$ $=$ $2\pi/\sqrt[3]{\pi R^2L}$ as a long-wavelength momentum cutoff for the system, and use atomic radii \cite{Slater1964} as a measure for maximal momentum cutoff which is reasonable for we can not probe atom's internal structure in our quantum optical treatment.\ For a needle-like cylinder where $H\gg A$, we have a redshift of several kHz that can be observable in conventional experimental setup.\ A blue shift is also shown in Fig. \ref{fig3} for a disk-like geometry ($A\lesssim H$) which is due to the counteracting term from system energy cutoff in Eq. (\ref{shift}).
\begin{figure}[t]
\centering
\includegraphics[width=12cm,height=5cm]{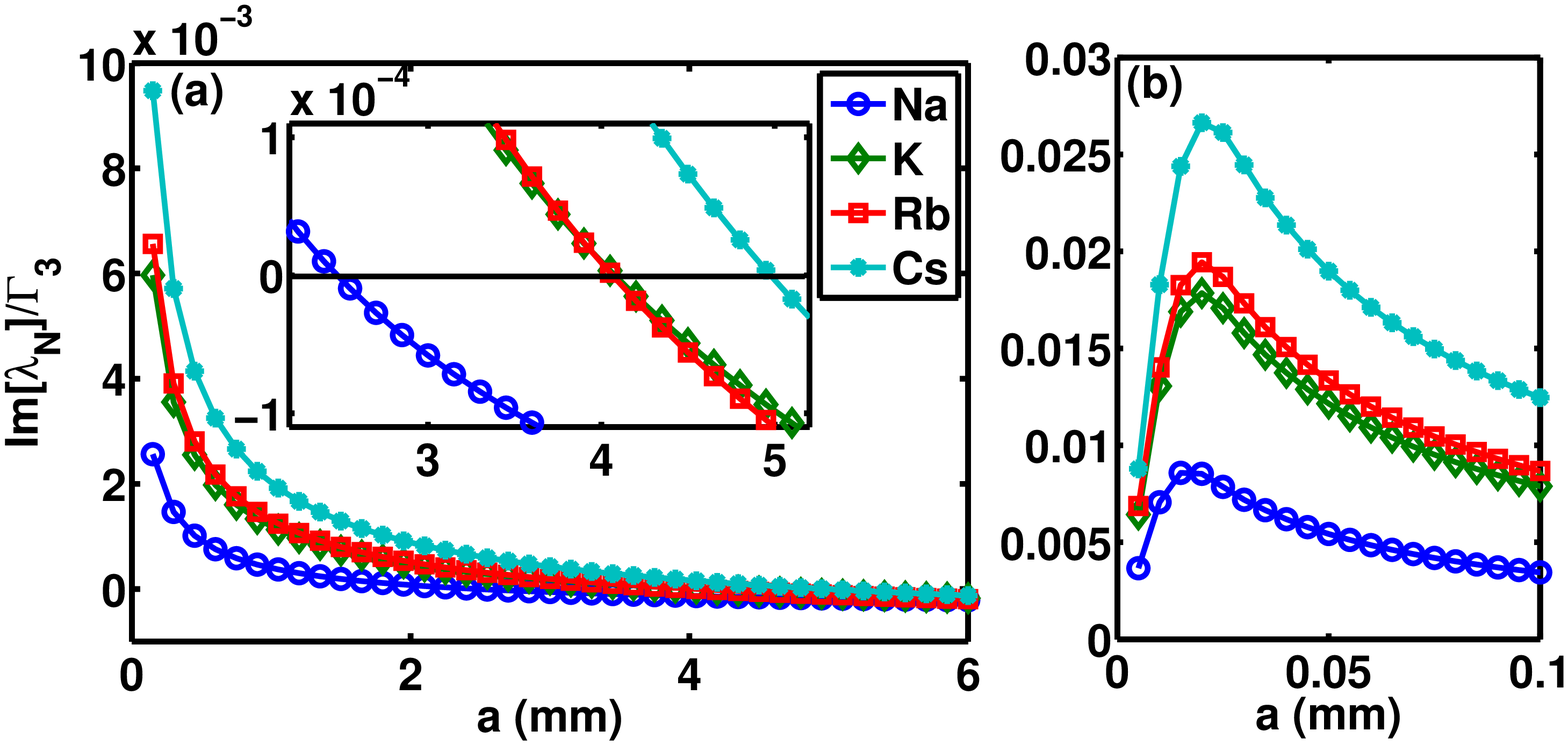}
\caption{(Color online) Collective Lamb shift of alkali metallic atoms as a dependence of radius $a$.\ The inset in (a) shows the crossing of red-blue shifts for different alkali metals, and short radius dependence is shown in (b).\ The height is chosen as $h$ $=$ $3$ mm.\ The atomic density and idler photon transition are the same as in Fig \ref{fig2}.}\label{fig4}
\end{figure}

In Fig. \ref{fig4}, we show the CLS as a dependence of radius, and demonstrate the crossing points of red-blue shifts for other alkali metals in the subplot (a).\ Due to different D1 transition wavelengths of alkali metals, sodium atoms reach the zero frequency shift at a smaller radius compared to rubidium and cesium ones.\ By scanning through the geometrical dependence and comparing with different atomic species, we may determine the system energy cutoff.\ The introduction of energy cutoff means lacking knowledge of the system momentum distribution as in the calculation of Eq. (\ref{cutoff}).\ If the experiment shows a redshift as radius varies, it means a unnecessary introduction of the short-wavelength energy cutoff, which suggests that CLS is only relevant to the macroscopic ensemble length scales ($H$ and $A$ in our case).

Note that when $H$ or $A$ approaches the condition when $N\rightarrow 1$, CLS is vanishing as can be seen in the general form of Eq. (\ref{f1}) where $N\bar{\mu}(k)\rightarrow 0$.\ In Fig. \ref{fig4}(b), we specifically show a short radius dependence where there is a sharp increase of CLS.\ The maximum appears at about $a$ $\approx$ $20$ $\mu$m, which can be significant if we have an even higher atomic density.\ An estimated of frequency shift of MHz is in the reach in the D1 transition of Rb atoms for atomic density $\rho\sim$ $10^{12}$ cm$^{-3}$.\ We note that similar red to blue shifts in CLS is demonstrated in an ellipsoid geometry \cite{Friedberg2010}.
\section{Conclusion}
We investigate the superradiance and cooperative Lamb shift (CLS) of correlated two-photon emissions in a diamond-type atomic ensemble.\ The photon pair is generated from the cascade transitions through four-wave mixing of two classical excitation fields.\ The superradiant idler photon is well described by one of the eigenvalues of coupling matrix constructed by the phased symmetrical state.\ This symmetrical state is significantly relevant when the number of atoms is large.\ We specifically analyze the geometrical dependence of CLS on a cylindrical atomic ensemble.\ We find a redshift in a needle-like geometry, and suggest it could be blue-shifted in a disk-like one.\ We further demonstrate a comparison of the CLS for different alkali metals, and propose to determine the system energy cutoff in experiments.\ 

We can take advantage of manipulating system's geometry and density to control the CLS, which enables robust frequency qubits in quantum network.\ In the same spirits of using dual-species matter and light-frequency qubits from a $^{85}$Rb-$^{87}$Rb isotopic mixture \cite{Lan2007}, we may prepare atomic ensembles with two different geometries or densities that generate entangled photon-pair states $(\hat{a}_{s,1}^\dagger\hat{a}_{i,1}^\dagger + \hat{a}_{s,2}^\dagger\hat{a}_{i,2}^\dagger)|{\rm vac}\rangle$ assuming the same excitation processes, where we denote (1,2) for two different frequencies and (s,i) for signal and idler bosonic fields $\hat{a}_{s,i}$ respectively.\ We may construct multimode frequency qubits without acquiring multiple atomic species, and control the qubits via optical means of atom traps which can be dynamical and efficient.\ The multimode frequency qubits may offer a speedup of entanglement generation as in temporal modes of rare-earth-metal ions \cite{Simon2007, Afzelius2009}, spatial quantum registers \cite{Tordrup2008, Grodecka-Grad2012}, and multiplexing quantum repeaters \cite{Collins2007}.
\section*{ACKNOWLEDGMENTS}
This work is supported by MOST (Ministry of Science and Technology) grant in Taiwan, and we thank T. A. B. Kennedy for the initiation and guidance of the work.
\appendix
\section{Interaction energy}
In this section, we derive the self-interaction energy shift for single and many two-level atoms interacting with a single photon.\ We demonstrate that a correct formulation of the interaction energy, including Lamb shift and CLS, is derived when non-RWA terms are included in the Hamiltonian.\ These non-RWA terms indicate the energy nonconserving process.\ It involves virtual excitations which are unphysical but crucial in determining the frequency shift.\ 

In interaction picture for single-atom light-matter interactions, we express the Hamiltonian as
\begin{align}
V&=\hbar\sum_\mathbf{k} g_\mathbf{k}\{|1\rangle \langle
0|\hat{a}_\mathbf{k} e^{-i(\omega-\omega_1)t}+|1\rangle \langle 0|\hat{a}^\dagger_\mathbf{k} e^{i(\omega+\omega_1)t} \nonumber\\&+|0\rangle \langle 1|\hat{a}_\mathbf{k}
e^{-i(\omega+\omega_1)t}+|0\rangle \langle 1|\hat{a}^\dagger_\mathbf{k} e^{i(\omega-\omega_1)t}\},
\end{align}
where $|0\rangle$ and $|1\rangle$ are ground and excited states respectively, and high frequency parts $e^{\pm i(\omega+\omega_1)t}$ are non-RWA terms.\ If the atom is in the excited state, it may decay to the ground state with an emitted single photon, whereas the atom in the ground state may be excited also by emitting a photon from non-RWA terms.\ The state function can be expressed as
\begin{align}
|\psi\rangle&=C_0(t)|0\rangle+C_1(t)|1\rangle+\sum_\mathbf{k} C_{1,\mathbf{k}}(t)|1,1_{\mathbf{k}}\rangle+\sum_\mathbf{k} C_{0,\mathbf{k}}(t)|0,1_{\mathbf{k}}\rangle.
\end{align}
The coupled equations from Schr\"{o}dinger equation are
\begin{align}
\dot{C}_0&=-i\sum_\mathbf{k} g_\mathbf{k} e^{-i(\omega+\omega_1)t}C_{1,\mathbf{k}},\nonumber\\
\dot{C}_1&=-i\sum_\mathbf{k} g_\mathbf{k} e^{-i(\omega-\omega_1)t}C_{0,\mathbf{k}},\nonumber\\
\dot{C}_{1,\mathbf{k}}&=-i g_\mathbf{k} e^{i(\omega+\omega_1)t}C_0,\nonumber\\
\dot{C}_{0,\mathbf{k}}&=-i g_\mathbf{k} e^{i(\omega-\omega_1)t}C_1.
\end{align}
Solving for the probability amplitudes of atomic energy levels, we have
\begin{align}
\dot{C}_0&=-\sum_\mathbf{k} |g_\mathbf{k}|^2 e^{-i(\omega+\omega_1)t}\int^t e^{i(\omega+\omega_1)t'}C_0(t')dt',\nonumber\\
\dot{C}_1&=-\sum_\mathbf{k} |g_\mathbf{k}|^2 e^{-i(\omega-\omega_1)t}\int^t e^{i(\omega-\omega_1)t'}C_1(t')dt'.
\end{align}

The summation of spontaneous emission modes in the above for the ground (sign $+$) and the excited (sign $-$) states becomes
\begin{align}
\int d\omega \frac{\Gamma}{2\pi}\left[\pi\delta(\omega\pm \omega_1)-i{\rm P.V.}(\omega\pm \omega_1)^{-1}\right],\nonumber
\end{align}
where the part of delta function simply represents the spontaneous decay rate, $\Gamma$ $\equiv$ $\frac{|d|^2\omega_1^3}{(3\pi\hbar\epsilon_0c^3)}$, with the dipole moment $d$ for the transition.\ P.V. denotes principal value.\ Moreover the Lamb shift can be identified as the difference of two level shifts,
\begin{align}
&\int d\omega\frac{\Gamma}{2\pi}\left[{\rm P.V.}(\omega-\omega_1)^{-1}-{\rm P.V.}(\omega+\omega_1)^{-1}\right]
\equiv\Omega^-_{\alpha\alpha}-\Omega^+_{\alpha\alpha}=\Omega.\nonumber
\end{align}

Now for many atoms interacting with a single photon including non-RWA terms, the Hamiltonian becomes
\begin{align}
V&=\hbar\sum_{\mu,\mathbf{k}} g_\mathbf{k}\Big\{|1\rangle_\mu \langle 0|\hat{a}_\mathbf{k} e^{i\mathbf{k}\cdot\mathbf{r}_\mu}e^{-i(\omega-\omega_1)t}
+|1\rangle_\mu \langle 0|\hat{a}^\dagger_\mathbf{k} e^{-i\mathbf{k}\cdot\mathbf{r}_\mu}e^{i(\omega+\omega_1)t}
\nonumber\\&+|0\rangle_\mu \langle 1|\hat{a}_\mathbf{k} e^{i\mathbf{k}\cdot\mathbf{r}_\mu} e^{-i(\omega+\omega_1)t}
+|0\rangle_\mu \langle 1|\hat{a}^\dagger_\mathbf{k} e^{-i\mathbf{k}\cdot\mathbf{r}_\mu}e^{i(\omega-\omega_1)t}\Big\},
\end{align}
where $\mu$ denotes the atomic indices.\ The state function can be written as
\begin{align}
|\psi\rangle&=C_0(t)|0\rangle+\sum_{\mu,\mathbf{k}} C_{1,\mathbf{k}}^\mu(t)|1_\mu,1_{\mathbf{k}}\rangle+\sum_\mu C_1^\mu(t)|1_\mu\rangle\nonumber\\
&+\sum_\mathbf{k} C_{0,\mathbf{k}}(t)|0,1_{\mathbf{k}}\rangle+\sum_{\mu,\mathbf{k}}\sum_{\nu\neq\mu}C_{2,\mathbf{k}}^{\mu\nu}(t)|1_\mu,1_\nu,1_{\mathbf{k}}\rangle,
\end{align}
where the last unphysical state (two atomic excitations with one photon present) is introduced to couple back to the singly-excited state \cite{Morawitz1973}.\ The coupled equations of motion for the probability amplitudes are
\begin{align}
\dot{C}_0&=-i\sum_{\mathbf{k},\mu} g_\mathbf{k} e^{-i(\omega+\omega_1)t}e^{i\mathbf{k}\cdot\mathbf{r}_\mu}C_{1,\mathbf{k}}^\mu,\label{C0}\\
\dot{C}_1^\mu&=-i\sum_k g_\mathbf{k} e^{-i(\omega-\omega_1)t}C_{0,\mathbf{k}}-i\sum_{\mathbf{k}}\sum_{\nu\neq\mu}g_\mathbf{k} e^{-i(\omega+\omega_1)t}e^{i\mathbf{k}\cdot\mathbf{r}_\nu}\left[C_{2,\mathbf{k}}^{\mu\nu}+(\mu\leftrightarrow\nu)\right],\label{C1}\\
\dot{C}_{1,\mathbf{k}}^\mu&=-i g_\mathbf{k} e^{i(\omega+\omega_1)t}e^{-i\mathbf{k}\cdot\mathbf{r}_\mu}C_0,\label{C1k}\\
\dot{C}_{0,\mathbf{k}}&=-i \sum_\mu g_\mathbf{k} e^{i(\omega-\omega_1)t}e^{-i\mathbf{k}\cdot\mathbf{r}_\mu}C_1^\mu,\label{C0k}\\
\dot{C}_{2,\mathbf{k}}^{\mu\nu}&= -i g_\mathbf{k} e^{i(\omega+\omega_1)t}e^{-i\mathbf{k}\cdot\mathbf{r}_\nu}C_1^\mu,\label{C2k}
\end{align}
where again the ground state energy level has a Lamb shift contribution similar to single atomic one except that an extra factor of $N$, the number of atoms, appears when substituting Eq. (\ref{C1k}) into Eq. (\ref{C0}), which shows its intrinsic, not collective, property by vacuum fluctuations.\ 

Carefully calculating the contribution from nonconserving energy states in the excited state by substituting Eq. (\ref{C2k}) into Eq. (\ref{C1}), we have
\begin{align}
&-i\sum_{\mathbf{k}}\sum_{\nu\neq\mu}g_\mathbf{k} e^{-i(\omega+\omega_1)t}e^{i\mathbf{k}\cdot\mathbf{r}_\nu}\left[C_{2,\mathbf{k}}^{\mu\nu}+(\mu\leftrightarrow\nu)\right]\nonumber\\
&=-\sum_\mathbf{k}|g_\mathbf{k}|^2e^{-i(\omega+\omega_1)t}\int^tdt'e^{-i(\omega+\omega_1)t'}\left[(N-1)C_1^\mu(t')+\sum_{\nu\neq\mu}
C_1^\nu(t')e^{i\mathbf{k}\cdot(\mathbf{r}_\nu-\mathbf{r}_\mu)}\right]\label{C1_2},
\end{align}
where we can combine the first term in the above with the Lamb shift contribution from the ground state (a factor of $N$), and we deduce the positive frequency part of Lamb shift $\Omega_{\alpha\alpha}^+$.\ By substituting Eq. (\ref{C0k}) into Eq. (\ref{C1}) for the energy conserving term and picking out the atomic index $\mu$, we deduce the other part of Lamb shift $\Omega_{\alpha\alpha}^-$.\ Again the Lamb shift is derived as $\Omega\equiv\Omega^-_{\alpha\alpha}-\Omega^+_{\alpha\alpha}$ which is the same as single atom case.

The second term in Eq. (\ref{C1_2}) is the dipole-dipole interaction energy from energy nonconserving terms.\ Along with the energy conserving terms by substituting Eq. (\ref{C0k}) into Eq. (\ref{C1}) and picking out the atomic indices other than $\mu$, we deduce the collective Lamb shift and spontaneous decay rate as
\begin{align}
&-\sum_\mathbf{k}|g_\mathbf{k}|^2\bigg\{e^{-i(\omega-\omega_1)t}\int^tdt'e^{i(w-w_1)t'}\sum_{\nu\neq\mu}C_1^\nu(t')e^{i\mathbf{k}\cdot(\mathbf{r}_\mu-\mathbf{r}_\nu)}\nonumber\\
&+e^{-i(\omega+\omega_1)t}\int^tdt'e^{-i(\omega+\omega_1)t'}\sum_{\nu\neq\mu} C_1^\nu(t')e^{i\mathbf{k}\cdot(\mathbf{r}_\nu-\mathbf{r}_\mu)}\bigg\}\nonumber\\
&=-\sum_{\nu\neq\mu} C^\nu_1(t)\left[\frac{\Gamma}{2}F_{\mu\nu}(\xi)-i(\Omega_{\mu\nu}^-(\xi)+\Omega_{\nu\mu}^+(\xi))\right],
\end{align}
where the property of frequency shift, $\Omega^\pm_{\alpha\beta}$ $=$ $\Omega^\pm_{\beta\alpha}$, is used, and $\xi$ $=$ $|\mathbf{k}| r_{\mu\nu}$, $r_{\mu\nu}$ $=$ $|\mathbf{r}_\mu-\mathbf{r}_\nu|$ with the transition wave vector $|\mathbf{k}|$.\ Here we show that the treatment of Schr\"{o}dinger's equation is equivalent to Heisenberg's picture that $F_{\alpha,\beta}$ and $G_{\alpha,\beta}$ are defined as \cite{Lehmberg1970}
\begin{align}
F_{\alpha,\beta}(\xi)&\equiv
\frac{3}{2}\bigg\{\left[1-(\hat{d}\cdot\hat{r}_{\alpha\beta})^2\right]\frac{\sin\xi}{\xi}+\left[1-3(\hat{d}\cdot\hat{r}_{\alpha\beta})^2\right]
\left(\frac{\cos\xi}{\xi^2}-\frac{\sin\xi}{\xi^3}\right)\bigg\},\\
G_{\alpha,\beta}(\xi)&\equiv\Omega_{\alpha\beta}/\Gamma_3\equiv-(\Omega^-_{\alpha\beta}+\Omega^+_{\alpha\beta})/\Gamma_3,\nonumber\\
&\equiv\frac{3}{4}\bigg\{-\Big[1-(\hat{d}\cdot\hat{r}_{\alpha\beta})^2\Big]\frac{\cos\xi}{\xi}+\Big[1-3(\hat{d}\cdot\hat{r}_{\alpha\beta})^2\Big]
\left(\frac{\sin\xi}{\xi^2}+\frac{\cos\xi}{\xi^3}\right)\bigg\},~{\rm for}~\alpha\neq\beta,
\end{align}
where $\hat{d}$ is the unit direction of electric dipole.\ The dependence of $|\xi|^3$ in the above earns the name of dipole-dipole interaction which is induced by the common light-matter interaction and radiation reaction.
\section{Adiabatic approximation of excitation process}
In this section, we derive the probability amplitudes of the excitation process in the adiabatic approximation.\ Firstly, we substitute the signal photon state of Eq. (\ref{signal_C}) into Eq. (\ref{eqn3}), and the summation of signal field modes in Eq. (\ref{eqn3}) becomes
\begin{align}
&\sum_{\mathbf{k}_s,\lambda_s}|g_s|^2|\epsilon^*_{\mathbf{k}_s,\lambda_s}\cdot\hat{d}_s|^2\int_{0}^tdt'e^{i(\omega_{\mathbf{k}_s}-\omega_{23}-\Delta_2)(t'-t)}
\sum_{l,m}e^{i\mathbf{k}_s\cdot(\mathbf{r}_\mu-\mathbf{r}_m)}\hat{U}_{\mu l}e^{\lambda_l(t-t')}\hat{U}^{-1}_{lm}B_m(t'),\nonumber\\
&\approx\sum_{l,m}[\frac{\Gamma_2}{2}F_{\mu m}(\xi')-i\Omega_{\mu m}^-(\xi')]\hat{U}_{\mu l}\hat{U}^{-1}_{lm}B_m(t),\nonumber\\
&=\frac{\Gamma_2}{2}B_\mu(t)+{\rm Lamb~ shift~ term},
\end{align}
where $\xi'$ $=$ $(\omega_{23}+\Delta_2)|\mathbf{r}_\mu-\mathbf{r}_m|/c$.\ The summation of the signal field modes is determined by the fast oscillating exponential factor of optical frequency, which is valid for the eigenvalues $\lambda_l$ $\ll$ $\omega_{23}$ $+$ $\Delta_2$.\ In the last step of derivation, we may absorb Lamb shift into the signal transition frequency, and the above result indicates that the upper excited state radiates in a single atomic decay rate.

In the limit of large detunings,
\begin{eqnarray}
|\Delta_1|,|\Delta_2|\gg\Omega_a,\Omega_b,\Gamma_2,\nonumber
\end{eqnarray}
we can solve the coupled equations of motion by adiabatically eliminating the intermediate and upper excited states in the excitation process.\ In Eqs. (\ref{eqn2}) and (\ref{eqn3}), we use integration by parts to express probability amplitudes in the first order of $1/\Delta_1$.\ Note that we allow time-varying excitation fields, and let $\tilde{A}_\mu(t)$$\equiv$ $e^{-i\mathbf{k}_a\cdot\mathbf{r}_\mu}A_\mu(t)$, $\tilde{B}_\mu(t)$$\equiv$$e^{-i(\mathbf{k}_a+\mathbf{k}_b)\cdot\mathbf{r}_\mu}B_\mu(t)$, we have
\begin{align}
\tilde{A}_\mu(t)&=e^{i\Delta_1 t}\Big[\frac{i}{2}\int_{-\infty}^t e^{-i\Delta_1 t'}\Omega_a(t')\mathcal{E}(t')dt'+\frac{i}{2}\int_{-\infty}^t e^{-i\Delta_1 t'}\Omega_b^*(t')\tilde{B}_\mu(t')dt'\Big],\nonumber\\
&=-\frac{\Omega_a(t)\mathcal{E}(t)}{2\Delta_1}-\frac{\Omega_b^*(t)\tilde{B}_\mu(t)}{2\Delta_1}+\mathcal{O}(\frac{1}{\Delta_1^2}),
\end{align}
\begin{align}\tilde{B}_\mu(t)&=e^{i(\Delta_2+i\Gamma_2/2)t}\Big[\frac{i}{2}\int_{-\infty}^t e^{-i(\Delta_2+i\Gamma_2/2)t'}\Omega_b(t')\tilde{A}_\mu(t')dt'\Big],\nonumber\\
&=-\frac{\Omega_b(t)\tilde{A}_\mu(t)}{2(\Delta_2+i\Gamma_2/2)}+\mathcal{O}(\frac{1}{\Delta_2^2}),
\end{align}
where the initial conditions $B({-\infty})$$=$$A({-\infty})=0$ are applied.\ The adiabatic approximation requires the driving pulses to be smoothly turned on.\ Therefore in the first order of adiabatic approximation, we derive
\begin{align}
\tilde{A}_\mu(t)&\approx-\frac{\Omega_a(t)}{2\Delta_1}\mathcal{E}(t),\\
\mathcal{E}(t)&\approx 1,\\
\tilde{B}_\mu(t)&\approx\frac{\Omega_a(t)\Omega_b(t)}{4\Delta_1\Delta_2}\equiv b(t).
\end{align}

The above results show that the probability amplitudes develop by following the driving fields, and the ground state is approximately unity in the limit of large detunings.\ The AC Stark shift in the ground state can be ignored if $N\int_{-\infty}^t|\Omega_a(t')|^2dt'$$\ll$$\Delta_1$ which is also required for the validity of assumption of single excitation in our scheme.

\end{document}